\documentclass[12pt]{article}

\usepackage{amsmath,amsthm,amsfonts, latexsym}
\usepackage{graphicx}

\begin{document}

\begin{center}

\bigskip
{\Large Quantum measurements and finite geometry}\\

\bigskip

William K.~Wootters\\

\bigskip

{\small{\sl

Department of Physics, Williams College, Williamstown, 
MA 01267, USA }}\vspace{3cm}

\end{center}
\subsection*{\centering Abstract}
{A complete set of mutually unbiased bases
for a Hilbert space of dimension $N$ is analogous in some respects
to a certain finite geometric
structure, namely, an affine plane.  Another kind of 
quantum measurement, known as a symmetric informationally complete
positive-operator-valued measure, is, remarkably, also analogous to
an affine plane, but with the roles of points and lines interchanged.
In this paper I present these analogies and ask whether they shed any 
light on the existence or non-existence of such symmetric quantum measurements
for a general quantum system with a finite-dimensional state space.}

\vfill

PACS numbers: 03.65.Ta, 03.65.Wj, 02.10.Ox

\newpage

\section{Introduction: Mutually unbiased measurements}

I have known Asher Peres since 1979, when he was visiting
John Wheeler at the University of Texas at Austin as I was 
finishing my graduate studies there.  In the years since then
our collaborations on various problems
in quantum mechanics have been
among the most enjoyable episodes of my career. 
One such 
collaboration took place in 1989 at the Santa Fe
Institute.  Asher raised the interesting question 
whether a joint measurement on a composite system could
ever discriminate among product states better than 
a series of separate measurements.  Our efforts towards
answering this question were
fueled in part, and were made much more interesting, 
by the interaction between our 
diametrically opposite intuitions
on the matter.  In those days I did not fully 
appreciate the depth of Asher's physical intuition
and was confident that I could prove him wrong. 
It was fun to try, and 
I learned a lot by trying, and I am grateful 
for the education.
For this and many other reasons, it is a pleasure
to dedicate this
paper to Asher on the occasion of his seventieth birthday.

In Austin in 1979 
Asher and I did not
collaborate on any paper, but we did 
discuss a few physics problems.  One of these was the problem of mutually 
unbiased measurements which is the starting point for the present 
article.  

For a system with an $N$-dimensional state space, a general 
mixed state is specified by $N^2 - 1$ real parameters.
Suppose we are given many copies of such a system and are 
trying to learn the values of these parameters.  If we perform a
fixed non-degenerate orthogonal measurement
on each of many copies,
we will eventually obtain $N-1$ independent real parameters, namely,
the probabilities of $N-1$ of the outcomes of our measurement.  (The
last probability is not independent since the probabilities must sum
to unity.)  By making a different orthogonal measurement on a different
set of copies of the system, we can hope to gain another $N-1$
real parameters, independent of the first set.  
Thus if we want to obtain all the 
parameters that define the quantum
state, and if we restrict ourselves to orthogonal measurements,
the minimum number of distinct
measurements we will need is $(N^2-1)/(N-1) = N+1$.  For example,
if the system in question is the spin of a spin-1/2 particle,
with $N=2$, we need at least three orthogonal measurements in order to
supply enough data to reconstruct the density matrix.  It is
natural to choose measurements corresponding to
three perpendicular spatial axes: up vs down, right vs left,
and in vs out.  From the perspective of minimizing the effects
of statistical fluctuations, these three are ideal in that they
are as different from each other as possible; each one
provides information that is maximally independent of the
information provided by the others.  

The relevant relationship that these three measurements
share is ``mutual unbiasedness'': each eigenstate of any one of them
is an equal-magnitude superposition of the eigenstates of any of the
others.  In $N$ dimensions, we say that two orthonormal bases
$\{|v_1\rangle, \ldots, |v_N\rangle\}$ and 
$\{|w_1\rangle, \ldots, |w_N\rangle\}$ are mutually unbiased
if $|\langle v_i|w_j\rangle|^2 = 1/N$ for each $i$ and $j$.
Because of the state-determination problem described above, it is
of some interest to find a set of $N+1$ mutually 
unbiased bases for an $N$-dimensional state space---we will
refer to such a set as a {\em complete} set of mutually
unbiased bases.  Indeed,
the problem is more interesting now than it was in the days when Asher
and I were discussing it in Austin.  For example,
mutually unbiased bases are relevant nowadays for
quantum cryptography.  The original quantum cryptographic schemes
used just two mutually unbiased bases in two dimensions
\cite{Wiesner,BB84}, but
a few years ago Asher, working with Helle Bechmann-Pasquinucci,
proposed more general schemes based on multiple unbiased bases
in higher dimensions \cite{Bechmann}.  

Is it possible to find $N+1$ mutually unbiased bases
in $N$ dimensions?  The answer is yes if $N$ is prime---Ivanovic
\cite{Ivanovic} constructed such bases in 1981---and 
the answer is again
yes if $N$ is any power of a 
prime 
\cite{WF,Calderbank,Lawrence,Bandy,Chaturvedi,Pittinger,
Zauner, Klapp}.  
Remarkably, though, the answer is
not known for any other values of $N$, not even for $N=6$.  We do know,
however, that for {\em any} value of $N$ the number of mutually unbiased bases
cannot exceed $N+1$ \cite{Ivanovic, Delsarte}.  

In this paper I would like to explore an
analogy, noted a few years ago by Zauner \cite{Zauner}
and more recently by Klappenecker and
R\"otteler \cite{Klapp}, between the problem of finding mutually unbiased bases
and an intriguingly similar
problem in combinatorics.  It is the well-known
problem of finding what are called mutually orthogonal Latin squares.  
The special case of a {\em complete}
set of $N+1$ mutually unbiased bases turns out to be analogous
to a finite geometric structure known as an 
affine plane.  This special case is the subject of a recent
conjecture
of Saniga {\em et al.} \cite{Saniga} that we will 
also consider in this paper.

Another special kind of quantum measurement that has attracted attention
lately is what is known as a
``symmetric informationally complete positive-operator-valued measure'' (SIC POVM) \cite{Renes}.
The second part of this paper discusses this sort of measurement and shows how it
too is analogous to a finite geometric structure.  Remarkably, the analogous
geometric structure is
{\em again} an affine plane, but with the roles of points and lines
interchanged.  After presenting these two analogies,
we will consider a number of open questions and briefly
discuss a connection with the foundations of quantum mechanics.  

Both mutually unbiased bases and SIC POVM's are special cases
of ``quantum designs'', which have been investigated
in the
dissertation of \hbox{Zauner} \cite{Zauner}. 
Indeed, Zauner has already noted not only the two analogies
on which I focus in this paper but others as well.  
What is probably most
novel about the work that I present here
is a connection---to be specified shortly---between
the affine-plane analogies and
the discrete phase space 
of Ref.~\cite{GHW}.  As we will see, this connection
suggests quantum analogues of  
the {\em points} of the finite geometries.

\section{Mutually orthogonal Latin squares}

The problem of finding mutually unbiased measurements is
similar in spirit to the following mathematical problem.
For any integer $N \geq 2$,
consider a collection of $N^2$ points, with no structure
except that the points are distinguishable from
each other.  Let a ``striation'' of this set 
be defined as any partitioning of the $N^2$ points
into $N$ disjoint subsets, called ``lines'', such that each line
consists of $N$ points.  Thus a striation defines a set of
$N$ lines that are parallel in the sense that no two of them have any
points in common.  Finally, let us call two striations
``mutually unbiased'' if each line in either striation has exactly
one point in common with each line in the other.  
A set of
four mutually unbiased striations for $N=3$ is shown in 
Fig.~1.  

\begin{figure}[h]
\centering
\includegraphics{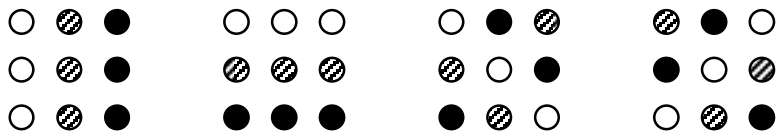}  
\caption{Four mutually unbiased striations of
nine points.}
\end{figure}  
 
I should point out that the above terminology is not standard; I am 
using it only to make the analogy clear.  In the usual formulation,
the $N^2$ points are imagined to be arranged in a square
(as in Fig.~1), and the
vertical and horizontal striations are assumed from the outset.  A ``Latin
square'' is then obtained by specifying a third striation in which each
line intersects each row and each column in exactly one point.
Two Latin squares are called orthogonal if their respective third
striations are mutually unbiased in the above sense.  Thus, for example,
finding a {\em pair} of orthogonal Latin squares amounts to finding 
{\em four} mutually unbiased striations according to the terminology
I am using here, because I am including the vertical and horizontal
striations in the count.

For a given value of $N$, it is natural to 
ask how many mutually unbiased striations
one can find.  Let $M(N)$ be the maximum possible
number of such striations.
This is a well-studied function,
but many questions about it remain unanswered.  The following
are some of the facts that are known \cite{Latinsquares}.

\begin{enumerate}
\item For any $N$, $M(N) \leq N+1$.  (The usual statement is that
the maximum number of mutually orthogonal Latin squares is no greater
than $N-1$.)
\item If $N$ is a power of a prime, this upper bound can be achieved;
that is, $M(N)$ is exactly equal to $N+1$.  Fig.~1 shows how the 
bound is achieved for $N=3$.
\item $M(6) = 3$. \cite{Tarry}
\item If $N-1$ or $N-2$ is divisible by four, and if $N$ 
is not the sum of
the squares of two integers, then $M(N)$
is {\em strictly less than} $N+1$. \cite{Bruck}
\item $M(10)$ is strictly less than 11. \cite{Lam}
\end{enumerate}

It is hard not the notice the parallels with the problem of mutually 
unbiased bases.  The first two items on the above list apply just as
well to the maximum number of 
mutually unbiased bases.  Regarding the third item, although
we do not know how many mutually unbiased bases one can find for
$N=6$, we can construct three of them, and there is some evidence that no more
than three can be found \cite{Zauner, Klapp, Archer}.  
Regarding the last two items, we simply 
do not have enough
evidence yet to know whether the analogous statements can be made about 
mutually unbiased bases, but it is not inconceivable that they can.  

One wonders, then, whether the two problems are in fact equivalent.
On the surface this may seem unlikely because the problems
seem to involve different constraints, 
but the question is worth exploring.
If the problems are equivalent, we should be able to find
mathematical entities in the unbiased basis problem that correspond
to the lines and points of the unbiased striation problem.  Clearly
we want each striation of the $N^2$ points to correspond to a basis 
for the complex vector space, and we want each line in a striation
to correspond to an element of the corresponding basis.
But what should a {\em point} correspond to?  

For the special case where $N$ is a power of a prime, there is
perhaps a natural interpretation of the points,
based on the discrete phase space of Ref.~\cite{GHW}.
This phase space is a two-dimensional vector space over the finite
field with $N$ elements, so it can be visualized as a
square array of $N^2$ points.  (There exists an $N$-element field
only if $N$ is a power of a prime.)  
The arithmetic
of the field is sufficient to define the concepts ``line'' and
``parallel lines'': a line is the
set of points satisfying a linear equation, and two lines are
parallel if they have the same slope but different intercepts.
One finds that the phase space can be partitioned into $N$ parallel lines
in exactly
$N+1$ ways.  Indeed, this 
construction is the basis of the standard proof that there
exists a complete set of mutually orthogonal Latin squares when 
$N$ is a power of a prime.  

The discrete phase space by itself does not yet provide a
quantum mechanical analogue of a ``point'' in the Latin square problem.
This analogue is supplied by the additional structure developed
in Ref.~\cite{GHW} for the purpose of representing quantum states
as functions on the discrete phase space.  This additional structure,
called a ``quantum net'', assigns to each line $\lambda$ of phase space
a pure quantum state, represented by a one-dimensional 
projection operator $P_\lambda$ in a space of $N$ dimensions.  
And it assigns
to each point $\alpha$ of phase space a Hermitian operator $A_\alpha$
on the same space
such that the following properties are satisfied:

\begin{enumerate}
\item Tr$(A_\alpha/N) = 1/N$
\item Tr$(A_\alpha/N)(A_\beta/N) = (1/N)\delta_{\alpha\beta}$
\item $\sum_{\alpha \in \lambda} (A_\alpha/N) = P_\lambda$
\end{enumerate}

\noindent It follows from these properties and from the geometry of 
the phase space that Tr$P_\lambda P_\nu = 0$ if $\lambda$
and $\nu$ are parallel lines, and that Tr$P_\lambda P_\nu = 1/N$
if $\lambda$ and $\nu$ are not parallel.  Since there are
$N+1$ sets of $N$ parallel lines, the projection operators
$P_\lambda$ thus define a complete set of $N+1$ mutually unbiased
bases for the state space. 

We see then that we can make 
the following correspondence between the Latin square
problem and the unbiased basis problem, when $N$ is a power of a 
prime.

\bigskip

\begin{center}

\begin{tabular}{c c c}

point $\alpha$ & $\leftrightarrow$ & operator $A_\alpha/N$ \\

 & & \\

line $\lambda$ & $\leftrightarrow$ & one-dimensional projection $P_\lambda$ \\

 & & \\

striation & $\leftrightarrow$ & orthonormal basis $\{P_{\lambda_1},\ldots,
P_{\lambda_N}\}$

\end{tabular}

\end{center}

\bigskip

\noindent The operation of composing $N$ points to make a line corresponds,
in the quantum setting, to taking the {\em sum} of the operators
$A_\alpha/N$ to obtain a one-dimensional projection operator
$P_\lambda$.  

Does this correspondence imply that when $N$ is a power
of a prime, the existence of $N+1$ mutually 
unbiased bases follows
immediately from the existence of $N+1$ mutually unbiased striations
in the Latin square problem?  No, because
it is by no means obvious how to construct the operators $A_\alpha$.
In Ref.~\cite{GHW} their construction depends on already having 
in hand a complete set of mutually unbiased bases, which are
obtained in a different way.  (The method used there to generate
these bases is essentially
the same as the one discovered independently 
by Pittinger and Rubin \cite{Pittinger}.)
Perhaps there is an alternative method
of constructing these operators such that the two problems can be
seen as equivalent.  At present I know of no such alternative construction.

It is worth
thinking further about the correspondence between geometric objects
in the Latin square problem and operators in the quantum problem.
In particular, one can make a connection between {\em cardinalities} of
sets of points and {\em traces} of operators
(see also Ref.~\cite{Zauner}, p. 23).  Let $M$ be an operator
such as $A_\alpha/N$ or $P_\lambda$, and let $S_M$ be the set of 
points that corresponds to that operator, if such a set exists.  
Let us look for relations of the form
\begin{equation}
|S_M|=k\,\hbox{Tr}\,M  \hspace{1cm}\hbox{and}\hspace{1cm} 
|S_{M_1} \cap S_{M_2}|=k\,\hbox{Tr}\,(M_1M_2), \label{keq}
\end{equation}
where $|\cdots |$ indicates the size of the set,
``$\cap$'' indicates the intersection of the two sets, and $k$
is a constant.  Following
the correspondence given above, we associate with the operator
$A_\alpha/N$ the set containing the single point $\alpha$, and with
the operator $P_\lambda$ the set containing the $N$ points of
the line $\lambda$.  Then the properties of the $A_\alpha$'s listed
above lead to the following equations:
\begin{enumerate}
\item $k/N = k\,$Tr$(A_\alpha/N) = |\{\alpha\}| = 1$
\item $(k/N)\delta_{\alpha\beta} = k\,$Tr$(A_\alpha/N)
(A_\beta/N) = |\{\alpha\}\cap\{\beta\}| = \delta_{\alpha\beta}$
\item $k = k\,$Tr$\,P_\lambda = |\{\hbox{points on the line }\lambda\}| = N$
\end{enumerate}
These conditions are indeed satisfied as long as we choose $k = N$.  

According to these rules, any operator that can be written
as the sum of some of the operators $A_\alpha/N$ corresponds to a set of points.
Other operators have no analogues as sets.  The identity, being the 
sum of all $N^2$ of the operators $A_\alpha/N$, corresponds to the
complete set of $N^2$ points in the Latin square problem.  Its
trace is $N$, corresponding to the fact that the cardinality of 
the set of all the points in the Latin square 
is $N^2$.  We will later consider a similar correspondence
between traces and cardinalities
in connection with our other quantum measurement problem.

Before moving on to that problem, it is interesting to say 
a little more about the geometry of orthogonal Latin squares.
When $N$ is such that one can find $N+1$ mutually unbiased
striations of $N^2$ points, one can show that 
the resulting geometric structure,
which includes a total of $N(N+1)$ lines, satisfies the following
simple rules (for $N\geq 2$):
\begin{enumerate}
\item Given any pair of points, there is exactly one line containing
both points.
\item Given any line $\lambda$ and any point $\alpha$ not lying on
$\lambda$, there is exactly one line through $\alpha$ that
is parallel to $\lambda$.
\item There exist three noncollinear points.
\end{enumerate}
Any set of points and lines satisfying these rules is called an
affine plane.  Every affine plane has $N^2$ points and $N(N+1)$ lines
for some value of $N$, and this value is called the {\em order} of the affine 
plane.  The $N(N+1)$ lines can be divided into $N+1$ sets of $N$
parallel lines, and two non-parallel lines always intersect in
exactly one point, so that any affine plane defines a complete set
of mutually orthogonal Latin squares.  
Thus the problem of finding a {\em complete}
set of $N+1$ mutually unbiased bases is analogous to finding
an affine plane of order $N$.  At present the only values of $N$
for which it is known that an affine plane of order $N$ exists
are the powers of primes.  And according to the facts we listed
earlier, there are some values of $N$ for which it is known that
an affine plane of order $N$ does {\em not} exist, {\em e.g.},
$N=6$ and $N=10$.  

Saniga {\em et al.} have recently conjectured that there exists
a complete set of mutually unbiased bases in $N$ dimensions
if and only if there exists an 
affine plane of order 
$N$ \cite{Saniga}.\footnote{Their paper is couched 
in terms of projective planes rather than affine planes,
but the essential content is the same either way.}  As we
have discussed, it is not yet clear whether our correspondence
between points and operators provides support for this 
conjecture, but it does provide a direction along which 
the question might be approached.  (See also the discussions 
of this issue by Zauner~\cite{Zauner} and Bengtsson~\cite{Bengtsson}.)

\section{Symmetric informationally complete positive-operator-valued measures}

Let us return to the scenario in which we are given many copies of
a quantum system and are trying to figure out what quantum state to
assign to the system.  In the Introduction we restricted our attention
to orthogonal measurements, and this restriction led us to the notion 
of a complete set of mutually unbiased measurements.  But 
other kinds of measurement
are certainly possible.  For a quantum system with an $N$-dimensional
state space, a positive-operator-valued measure (POVM)
is a set of positive operators $E_i$ such that $\sum_i E_i = I$,
where $I$ is the $N\times N$ identity.  A POVM represents a 
quantum measurement for which the probability of the $i$th outcome
is Tr$(\rho E_i)$, $\rho$ being the system's density matrix. 
Note that the operators $E_i$ need not be orthogonal to each
other; that is, Tr$E_iE_j$ need not equal zero when $i\neq j$. 
Most physicists did not know about POVMs when I met Asher,
and indeed his book, {\em Quantum Mechanics: Concepts and Methods},
is still unusual among quantum mechanics textbooks in explaining
or even mentioning POVMs \cite{Asherbook}.

Whereas the state-reconstruction scheme we discussed earlier required
several distinct orthogonal 
measurements, it is possible to get the same information
by means of a {\em single} POVM performed on many copies
of the system.  Since, 
as before, there are $N^2 - 1$ real parameters
to be determined, this single POVM would have to have at least $N^2$
outcomes, thus providing at least $N^2 - 1$ independent probabilities.
In the case of a spin-1/2 particle, a minimal POVM capable of extracting
the necessary parameters would have exactly four outcomes, and the
corresponding four probabilities will be maximally independent if
we make the operators $E_i$ in some sense 
as different from each other as possible.  The natural choice is
to let $E_i = (1/2)P_i$, where the one-dimensional projectors
$P_i$ correspond to four tetrahedrally related points on the 
Bloch sphere.  The operators $E_i$ in this case constitute
what is called a symmetric
informationally complete POVM, or SIC POVM \cite{Renes}.  

In $N$ dimensions, a SIC POVM is a set of $N^2$ operators of the
form $E_i = (1/N)P_i$, where the one-dimensional projectors $P_i$
satisfy the condition
\begin{equation}
\hbox{Tr}\,P_i P_j = \frac{1}{N+1} \hspace{1cm} i\neq j.
\end{equation}
One can show that such a set constitutes a POVM and is informationally
complete in the sense that the probabilities it provides are sufficient
to reconstruct any $N\times N$ density matrix.  

This approach to state-reconstruction leads to another mathematical
question about complex vector spaces: for a space of $N$ dimensions,
does there exist a SIC POVM?  Like the question about mutually 
unbiased bases, the answer is known only in certain cases.
Here is a summary of our current state of knowledge
\cite{Lemmens, Koldobsky, Zauner, Renes, Grassl}:

\begin{enumerate}

\item For $N = 2$, 3, 4, 5, 6, and 8, 
SIC POVM's exist and explicit expressions
for them are known.  (In the above discussion we essentially demonstrated
the existence of a SIC POVM for $N=2$.  The one for $N=4$ 
\cite{Zauner, Renes} is considerably 
less obvious!)

\item For every $N \leq 45$, there is good numerical evidence that 
a SIC POVM exists \cite{Renes}.  

\item There is no known {\em proof} that a SIC POVM exists for any value of $N$ other than those listed in item 1.

\end{enumerate}

Because of the numerical evidence, it is plausible that a SIC POVM
exists in every dimension, though if this is indeed the case, it is
remarkable that a proof of this existence is so
elusive.

\section{Affine planes through the looking glass}

Let us now try to construct a finite geometric problem analogous to
the problem of finding a SIC POVM.  As before, we will let
{\em lines} be the geometric objects that correspond to
one-dimensional projections $P$.  But how many points 
will lie on a line?  And how many points will there be
altogether?  Taking some guidance from our earlier discussion, 
let us assume relations of
the following
form between cardinalities of sets of points and traces
of operators:
\begin{equation}
|S_M|=k'\,\hbox{Tr}\,M  \hspace{1cm}\hbox{and}\hspace{1cm} 
|S_{M_1} \cap S_{M_2}|=k'\,\hbox{Tr}\,(M_1M_2), \label{k'eq}
\end{equation}
where the constant $k'$ is not necessarily equal to the $k$
of Eq.~(\ref{keq}).

For each line $\lambda$, the corresponding projection operator $P_\lambda$
has trace equal to 1.  Therefore, according to Eq.~(\ref{k'eq}), 
the number of points on this line should be $k'$.  By definition of
a SIC POVM, we also have that for two distinct lines
$\lambda$ and $\nu$, Tr$P_\lambda P_\nu = 1/(N+1)$, from which
it follows that $k'/(N+1)$ is the number of points in the intersection of
$\lambda$ and $\nu$.  This number must be an integer, so $k'$ must be
an integral multiple of $N+1$.  The simplest possibility, then, which also
has the pleasing feature that two lines intersect in exactly one point,
is to choose $k' = N+1$.  Thus, in our geometry each line will contain
$N+1$ points, and each pair of lines will intersect in one point.
How many points should there be altogether?  As in the Latin square problem,
let us assume that the set of all points corresponds to the identity operator.
Then Eq.~(\ref{k'eq}) tells us that the total number of points must be
$k'\,\hbox{Tr}\,I = N(N+1)$.  Finally, let us assume for the sake of
symmetry that each point lies on the same number of lines as each other
point.  
A simple counting argument then tells us that 
each point lies on exactly $N$ lines.

To find a geometric model of the SIC POVM problem, then, we are
looking for a geometry in which 
\begin{enumerate}
\item there are exactly $N(N+1)$ points
\item there are exactly $N^2$ lines
\item each line contains exactly $N+1$ points
\item each point lies on exactly $N$ lines
\item each pair of lines intersect in exactly one point
\end{enumerate}
It turns out that these conditions precisely describe 
an affine plane, which we have seen before, except that
the roles of points and lines have been reversed.  Thus we
already know something about the values of $N$ for which we
can find the kind of geometric structure we are looking for:
we can find such a structure 
when $N$ is a power of a prime, and there
are other values of $N$ (e.g., $N=6$ and $N=10$)
for which we know that no such structure
exists.  

To see how the Latin square problem turns into our current
problem when we interchange points and lines, let us consider
the case $N=2$.  We start with a square array of four points
as shown in Fig.~2(a), where we have also drawn the six lines
that define three mutually unbiased striations.  Replacing
each point with a line and vice versa, while maintaining the
coincidence relations between points and lines, we obtain the
structure shown in Fig.~2(b).  (The circle in the figure counts
as one of the lines.)  
Note that in this structure, the
six points can be grouped into three pairs such that the points in each 
pair are {\em not} connected by a line, just as in Fig.~2(a),
the six lines can be grouped into three pairs of lines
such that the lines in each pair have no point in common. 

\begin{figure}[h]
\centering
\includegraphics{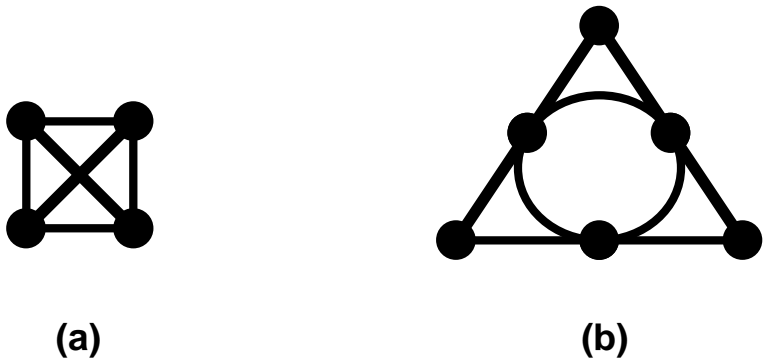}  
\caption{(a) The affine plane of order 2.
(b) The structure resulting from this affine
plane upon interchanging the
roles of points and lines.}
\end{figure}

Note that there is a limitation to the geometric
analogy for the SIC POVM problem.  
For certain small values of $N$ such as 6 and 10, 
we know that there does not exist
an affine plane of order $N$, and yet there does exist
a SIC POVM for $N=6$ \cite{Grassl}, 
and the numerical evidence cited
earlier strongly suggests that SIC POVM's exist for
all dimensions up to $N=45$.  Evidently there are ways of 
constructing SIC POVM's that have no particular relation to
affine planes.  

Nevertheless, let us continue with the analogy for those values
of $N$ for which affine planes exist, and try to associate a Hermitian
operator $B_\alpha$ with each point $\alpha$ of our geometry.
(In this discussion it helps to keep Fig.~2(b) in mind as
an example.)
As in the Latin square problem, we would like the sum of $B_\alpha$
over each line $\lambda$ to be the projection operator $P_\lambda$
associated with that line.  Our rule for relating traces to
cardinalities requires that for each point $\alpha$, Tr$B_\alpha
= 1/(N+1)$.  What about the traces of products $B_\alpha B_\beta$?
We need to choose these so that Tr$(P_\lambda P_\nu) = 1/(N+1)$
if $\lambda \neq \nu$ and Tr$(P_\lambda^2) = 1$.  One can show
that these two relations are guaranteed if we insist on
the following trace relations among the $B_\alpha$'s:
\begin{enumerate}
\item Tr$(B_\alpha^2) = \frac{N}{(N+1)^2}$.
\item Tr$(B_\alpha B_\beta) = \frac{1}{N(N+1)^2}$ if $\alpha\neq\beta$ and
$\alpha$ and $\beta$ share a line.
\item Tr$(B_\alpha B_\beta) = -\frac{1}{(N+1)^2}$ if $\alpha$ and $\beta$
do not share a line.
\end{enumerate}

For the affine plane of order $N$, can we find such a set of 
operators $B_\alpha$?  Here I answer the question in the simplest
case $N=2$, pictured in Fig.~2(b), and leave the question open for other values of $N$.
For $N=2$, we can indeed find such operators.  Let each $B_\alpha$
be of the form 
\begin{equation}
B_\alpha = \frac{I}{6} \pm \frac{\sigma}{2\sqrt{3}} ,
\end{equation}
where $\sigma$ is one of the three Pauli matrices.  The 
six operators $B_\alpha$ are to be assigned to the six
points of our geometry in such a way that operators
differing only in the sign of the Pauli matrix are assigned
to points that do not share a line.  One can verify then
that the $B_\alpha$'s satisfy the three conditions
listed above, as well as the normalization 
condition Tr$\,B_\alpha = 1/3$.

Will these ideas make it easier to find SIC POVM's, at least 
for those values of $N$ for which affine planes exist?   
It is not clear that they will, because there is no obvious
method of constructing operators $B_\alpha$ that satisfy 
the conditions given above.  (If one already {\em has} a 
SIC POVM, one can work backward from the projection operators
$P_\lambda$ to find such a set of $B_\alpha$'s, but in that case
the $B_\alpha$'s must not have been of much help.)  But perhaps such a method can
be found, at least for certain values of $N$.

\section{Discussion}

For two problems concerning 
quantum measurements---one dealing with mutually
unbiased bases and the other with SIC POVM's---we have found
analogues in finite geometry.  In each case I have
suggested (as have other authors) that a {\em line}
in the geometry is to be associated with a {\em pure state} in the quantum 
problem.  We have found that it also makes sense to associate
with each {\em point} of the geometry a Hermitian operator, such that
the sum of the operators corresponding to all the points on a given 
line defines the
quantum state that is to be associated with that line.

At present there is no evidence that these analogies will 
help us find either mutually unbiased bases or SIC POVM's.  But it
is conceivable that they may end up being a piece of the
puzzle.  In the SIC POVM problem, for example, 
if one can find a simple prescription for generating
the operator associated with each point, one would
be able to generate SIC POVM's for at least certain values of the
dimension of the state space. 

These analogies suggest a number of questions.  As we have seen, the geometric
structure analogous to a complete set of $N+1$ mutually unbiased bases is
{\em identical} to the geometric structure analogous to a SIC POVM,
except that the roles of points and lines are interchanged. 
Is there any sense in which the problem of finding $N+1$ mutually 
unbiased bases and the problem of finding a SIC POVM are likewise equivalent,
at least for those values of $N$
for which affine planes exist?  
Suppose, for example, that we are given a complete set of mutually unbiased
bases, which is a set of $N(N+1)$ state vectors (or one-dimensional
projection operators).  Can we regard these state vectors as the
``points'' from which we can construct $N^2$ ``lines'' that
are the elements of a SIC POVM?  Or, if we are given a set of
$N^2$ state vectors that define a SIC POVM, can we regard these
state vectors as the ``points'' of a square array, from which we can
construct $N+1$ sets of ``parallel lines'' that constitute $N+1$
mutually unbiased bases?  

Numerical evidence makes it reasonable to conjecture that SIC POVM's
exist in every dimension, whereas numerical searches have failed to
find a complete set of mutually unbiased bases even in six dimensions.
Does this mean that the suggestion of equivalence is misleading
and that these two problems are in fact entirely unrelated?
Does the geometric analogy, which definitely favors certain
values of $N$, have deep significance for the problem
of mutually unbiased measurements while being nothing but a red herring 
for the problem of SIC POVM's?  Or is it a red herring in both 
cases?  Presumably it is only a matter of
time, possibly a short time, before we know the answers to these questions.

All of the above questions are mathematical in nature.  But it is also interesting
to think about these geometric constructions from the perspective of
the foundations of quantum mechanics.  

Suppose that, for a system with
an $N$-dimensional state space, the {\em only} pure states available to
the system were the $N^2$ states defined by a SIC POVM.  Then we could
construct a hidden-variable model along the following lines.
Let there be exactly $N(N+1)$ hidden ``ontic'' states available to the system.
These are represented by the $N(N+1)$ points of our finite geometry.
The ``epistemic'' states, which we perceive as pure quantum states, 
are the $N^2$ lines in the geometry, each consisting
of $N+1$ ontic states.  When we assign one of these epistemic 
states to the system,
we do so (according to this model) because we do not 
know its ontic state, and the most we can
ever know about its ontic state is that it lies on a particular line.
(These ideas are very much in the spirit of the toy models of 
quantum mechanics recently proposed and analyzed by 
Spekkens \cite{Spekkens} and Hardy \cite{Hardy}.)
We now make a measurement on the state, supposing that the only
measurements available to us are yes-no measurements represented
by the one-dimensional projection operators that define our 
SIC POVM.
That is, the system is in some state $|\psi_i\rangle$, represented
by a line $\lambda_i$ in our geometry, and we are asking whether it will be found
to be in some other state $|\psi_j\rangle$, represented by a different
line $\lambda_j$.  To compute the probability of ``yes'', we simply count how many
points of $\lambda_i$ are also in $\lambda_j$.  The result is 
this: of
the $N+1$ points in $\lambda_i$, exactly one is in $\lambda_j$; so the 
probability of ``yes'' is $1/(N+1)$.  This is indeed the correct
quantum mechanical probability.  The agreement is not surprising
and is in fact guaranteed by our construction.
By making the traces of operators proportional to the
cardinalities of the corresponding sets, we ensured that the geometry 
would  
produce the correct probabilities. 

Can one use a similar construction to generate 
quantum mechanics itself rather than a limited model of quantum
mechanics?  By ``similar construction'' I mean that there is a set of
points representing the underlying (but hidden) 
ontic states, and that what
we call pure quantum states are represented by special subsets of
these ontic states.  Orthogonal states would be represented by
disjoint subsets, a complete orthogonal basis would be represented by
a partitioning of the whole set,
and probabilities would be computed from the
sizes of the intersections of subsets.  The answer 
is {\em no};
no such model can reproduce all the probabilities given by quantum mechanics.
A model of this sort would be a non-contextual hidden variable theory,
and such theories are ruled out by 
the Kochen-Specker theorem \cite{KS}.\footnote{Indeed, the kind of
model we are considering is a special case and can be ruled out 
in other ways as well.}
It is true that our
symmetric collection of $N^2$ states in $N$ dimensions can be accommodated
within such a model,
but in a typical proof of the Kochen-Specker theorem one
identifies some other collection of quantum states for which no such model
exists.  One of the most elegant proofs of the Kochen-Specker theorem,
based on a particularly symmetric set of 33 pure states in 
three dimensions, is due to Asher Peres \cite{PeresKS}.   

It would be interesting to find out how far one can go towards
mimicking quantum mechanics with a theory in which pure states
are represented as subsets of some larger set
of ontic states.  In $N$ dimensions, what is the greatest number of
vectors one can find such that the squares of their inner products
can be obtained from the sizes of the intersections of the corresponding subsets? 
If $N$ is a power of a prime, we can find at least $N(N+1)$ such vectors, namely,
the elements of all the mutually unbiased bases.  Perhaps
we can find more.    
Regardless of the result, however, we cannot go all the way.
There are aspects of quantum 
mechanics that accord with our classical intuition---the relationships
among a special collection of states can serve as an example---but the
theory as a whole, and the world to which it applies, are 
profoundly at odds with the framework of classical physics. 

\section*{Acknowledgements}

I would like to thank Robin Blume-Kohout and Robert Spekkens
for discussions about quantum measurements
and finite geometry, and 
Markus Grassl, Bengt Nagel, and Rasmus Hansen for helpful
comments on earlier versions of the paper.

\end{document}